\documentclass{aa}
\usepackage{txfonts}

\usepackage{graphicx, epsfig}

\usepackage{natbib}
\bibpunct{(}{)}{;}{a}{}{,} % to follow the A&A style

\usepackage{dcolumn}

\begin{document}

%Definitions to arrange columns at special characters (\usepackage{dcolumn}
%\newcolumntype{character which is used \begin{tabular}{l.d{4}..l} }[??]
%{D{Character in table}{Character for the dvi-file}{precision of numbers}} 
\newcolumntype{.}{D{.}{.}{2}}
\newcolumntype{d}[1]{D{,}{\,\pm\,}{#1}}

%%%%%%%%%%
% Header %
%%%%%%%%%%

% Title, author's and institutes
\title{Kinematic study of the blazar \object{S5 0716+714}}

\author{U. Bach\inst{1}\thanks{Current address: INAF - Osservatorio
Astronomico di Torino, Via Osservatorio 20, 10025 Pino Torinese, Italy}
	\and T.P. Krichbaum\inst{1}
	\and E. Ros\inst{1}
	\and S. Britzen\inst{1,2}
        \and W.W. Tian\inst{1,3}
	\and A. Kraus\inst{1}
	\and A. Witzel\inst{1}
	\and J.A. Zensus\inst{1}}
\authorrunning{U. Bach et al.} 
\institute{Max-Planck-Institut f\"ur Radioastronomie, Auf dem H\"ugel 69, 53121 Bonn, Germany 
	\and Landessternwarte, K\"onigstuhl 17, 69117 Heidelberg, Germany
        \and National Astronomical Observatories, CAS, 20A Datun, Road Chaoyang, Beijing 100012, PR China}
%Beijing Astronomical Observatory and Astrophysics Center of the
%
% First author's information
\offprints{U. Bach, \email{bach@to.astro.it}}
% Date 
\date{Received 5 March 2004; accepted 3 December 2004}

%%%%%%%%%%%%
% Abstract %
%%%%%%%%%%%%
\abstract{We present the results of a multi-frequency study
of the structural evolution of the VLBI jet in the BL\,Lac object
0716+714 over the last 10 years. We show VLBI images obtained at
5\,GHz, 8.4\,GHz, 15\,GHz and 22\,GHz. The milliarcsecond source
structure is best described by a one-sided core-dominated jet of $\sim
10$\,mas length. Embedded jet components move superluminally with
speeds ranging from $5\,c$ to $16\,c$ (assuming $z=0.3$). Such fast superluminal
motion is not typical of BL\,Lac objects, however it is still in the range of jet
speeds typically observed in quasars ($10\,c$ to $20\,c$). In 0716+714, younger
components that were ejected more recently seem to move 
systematically slower than the older components. This and a systematic
position angle variation of the inner (1\,mas) portion of the VLBI jet suggests
an at least partly geometric origin of the observed velocity variations. The
observed rapid motion and the derived Lorentz factors are discussed with regard
to the rapid {\bf I}ntra-{\bf D}ay {\bf V}ariability (IDV) and the $\gamma$-ray
observations, from which very high Doppler factors are inferred.
}
%
% Finish the title
\maketitle

%%%%%%%%%%%%%%%%%
% Begin Article %
%%%%%%%%%%%%%%%%%

\section{Introduction}
The S5 blazar 0716+714 is one of the most active BL\,Lac objects. It is
extremely variable on time-scales from hours to months at all observed
wavelengths from radio to X-rays. The redshift of 0716+714 is not yet
known. However, optical imaging of the underlying galaxy provides a
redshift estimate of $z \geq 0.3$ (\citealt{1996AJ....111.2187W}). A recent
X-ray study by \cite{2004A&A.....0....0K} suggests a much lower redshift of 
0.1, but since this needs further confirmation a redshift of 0.3 is used
throughout this paper. In the radio bands 0716+714 is an intraday variable (IDV)
source (\citealt{1986MitAG..65..239W}; \citealt{1987AJ.....94.1493H}). It
exhibits a flat radio spectrum at frequencies up to at least 350\,GHz.
\cite{1991ApJ...372L..71Q} found strongly correlated IDV between radio and
optical bands. This and the correlated variations between X-ray and optical and
the simultaneous variations between optical and
radio strongly suggest an intrinsic origin of the intraday variability
(\citealt{1996AJ....111.2187W,1995ARA&A..33..163W}). From the observed IDV at cm-wavelengths a
typical brightness temperature of $T_{\rm b} = 10^{15.5}$\,K to $10^{17}$\,K is
derived (\citealt{2003A&A...401..161K, 1993A&A...271..344W}). A Doppler factor
of 15 to 50 would be needed to bring these brightness temperatures down to the
inverse-Compton limit of $10^{12}$\,K. 

Very Long Baseline Interferometry (VLBI) studies spanning more than 20\,years
at cm-wavelengths show a core-dominated evolving jet extending several
10\,milliarcseconds to the north
(\citealt{1986A&A...168...17E,1987A&AS...67..121E, 1988A&A...206..245W}). The
VLBI jet is misaligned with respect to the VLA jet by $\sim 90\,^\circ$ (e.g.,
\citealt{1987MNRAS.228..203S}). In the literature the jet kinematics in
0716+714   are discussed controversially. There exist several kinematic
scenarios with motions ranging from 0.05\,mas\,yr$^{-1}$ to
1.1\,mas\,yr$^{-1}$. However, some of these earlier kinematic models are based
on only two or three epochs separated by several years, which sometimes could
have led to ambiguities in the identification of fast components
(\citealt{1987A&AS...67..121E,
1988A&A...206..245W,1992vob..conf..225S,1998A&A...333..445G,2004A&A.....0...0P}).
More recent  studies based on more data measured higher jet speeds
(\citealt{2001IAUS..205...96T, 2001ApJS..134..181J,2004ApJ...609..539K}), but the results were
still inconsistent between the two studies. Thus, it is at present not clear if
0716+714 is a superluminal source and and how fast the motion of the VLBI jet
components is.

In this paper, we present and discuss our results from a reanalysis of the last
10 years of VLBI data on 0716+714, obtained at frequencies between 5\,GHz and
22\,GHz. Our analysis includes 26 observing epochs listed in
Table~\ref{tab:obs}. Throughout this paper we will use a flat universe, with the
following parameters: A Hubble constant of $H_0=71$\,km\,s$^{-1}$\,Mpc$^{-1}$,
a pressureless matter content of $\Omega_{\rm m}=0.3$ and a cosmological constant of 
$\Omega_{\rm \lambda}=0.7$. With these constants an angular motion of 1\,mas/yr
corresponds to a speed of 18.8\,$c$ at $z=0.3$ (6.6\,$c$ at $z=0.1$).

In Sect.~\ref{sec:obse}, we briefly describe the observations and
the data reduction, the model fitting process and present the final
images. In Sect.~\ref{sec:discuss} we present the cross-identification
of the individual jet components and a new kinematic model for the jet in
0716+714. From this we derive the jet speed and orientation.
We summarise our results in Sect.~\ref{sec:conclusion}.

\begin{table}[htbp]
\caption{Observation log. Listed are the observing epoch, frequency $\nu$, total
flux density $S_{\rm tot}(\nu)$, beam size, beam position angle, peak flux
density $S_{\rm peak}$ and the lowest contour at 3\,$\sigma$ of the map in
 Figs.~\ref{fig:allmaps1}~\&~\ref{fig:allmaps2}.} 
\label{tab:obs}
%\small
\centering
\begin{tabular}{lrcccc}
\hline
Epoch & 
\multicolumn{1}{c}{$\nu$} & 
\multicolumn{1}{c}{$S_{\rm tot}$} & 
\multicolumn{1}{c}{Beam} &
\multicolumn{1}{c}{$S_{\rm peak}$} & 
\multicolumn{1}{c}{3\,$\sigma$}\vspace{1pt}\\
 & 
\multicolumn{1}{c}{[GHz]} & 
\multicolumn{1}{c}{[Jy]} & 
\multicolumn{1}{c}{[mas\,$\times$\,mas], [$^\circ$]} &
\multicolumn{1}{c}{$[\frac{Jy}{beam}]$} & 
\multicolumn{1}{c}{$[\frac{mJy}{beam}]$}\vspace{1pt}\\
\hline                                                                              %rms [Jy]
1992.73\,$^{a,(1)}$&  5.0 & 0.67 & $0.8\times1.0$,\hspace{9.5pt}   0 & 0.61 & 1.5 \\ %rms 0.00034
{\bf 1992.85}\,$^a$& 22.2 & 0.75 & $0.9\times2.7$,\hspace{0pt} $-55$ & 0.64 & 3.0 \\ %rms 0.00077
{\bf 1993.71}\,$^a$& 22.2 & 0.40 & $0.2\times0.2$,\hspace{5.5pt}  35 & 0.31 & 2.0 \\ %rms 0.00079
{\bf 1994.21}      &  8.4 & 0.32 & $0.8\times1.4$,\hspace{5.5pt}  80 & 0.26 & 0.7 \\ %rms 0.00020
{\bf 1994.21}      & 22.2 & 0.34 & $0.3\times0.5$,\hspace{5.5pt}  79 & 0.30 & 1.8 \\ %rms 0.00072
1994.67\,$^{(2)}$  & 15.3 & 0.46 & $0.5\times0.7$,\hspace{0pt} $-29$ & 0.40 & 1.8 \\ %rms 0.00056
1994.70\,$^{a,(1)}$&  5.0 & 0.36 & $0.8\times1.2$,\hspace{0pt} $-15$ & 0.29 & 2.7 \\ %rms 0.00070
1995.15\,$^{(3)}$  & 22.2 & 0.75 & $0.3\times0.5$,\hspace{4pt}  $-3$ & 0.71 & 4.0 \\ %rms 0.00200
1995.31\,$^{(3)}$  & 22.2 & 0.43 & $0.3\times0.4$,\hspace{5.5pt}  39 & 0.40 & 4.2 \\ %rms 0.00218
1995.47\,$^{(3)}$  & 22.2 & 0.20 & $0.3\times0.6$,\hspace{9.5pt}   7 & 0.18 & 3.0 \\ %rms 0.00135
{\bf 1995.65}      &  8.4 & 0.31 & $0.9\times1.0$,\hspace{5.5pt}  54 & 0.26 & 0.4 \\ %rms 0.00012
{\bf 1995.65}      & 22.2 & 0.34 & $0.3\times0.4$,\hspace{0pt} $-40$ & 0.28 & 1.8 \\ %rms 0.00082
1996.34\,$^{(3)}$  & 22.2 & 0.27 & $0.3\times0.9$,\hspace{5.5pt}  48 & 0.22 & 3.2 \\ %rms 0.00158
1996.53\,$^{(2)}$  & 15.3 & 0.26 & $0.5\times0.7$,\hspace{0pt} $-15$ & 0.22 & 0.7 \\ %rms 0.00022
1996.60\,$^{(3)}$  & 22.2 & 0.25 & $0.4\times0.4$,\hspace{9.5pt}   9 & 0.21 & 3.0 \\ %rms 0.00110
1996.63\,$^{a,(1)}$&  5.0 & 0.22 & $1.4\times2.2$,\hspace{5.5pt}  41 & 0.18 & 1.3 \\ %rms 0.00049
1996.82\,$^{(2)}$  & 15.3 & 0.26 & $0.4\times0.7$,\hspace{5.5pt}  21 & 0.25 & 2.0 \\ %rms 0.00076
1996.90\,$^{(3)}$  & 22.2 & 0.29 & $0.4\times0.5$,\hspace{0pt} $-35$ & 0.27 & 2.0 \\ %rms 0.00078
1997.03\,$^{(4)}$  &  8.4 & 0.21 & $1.1\times1.4$,\hspace{9.5pt}   7 & 0.18 & 0.0 \\ %rms 0.00177
1997.58\,$^{(3)}$  & 22.2 & 0.98 & $0.4\times0.5$,\hspace{0pt} $-33$ & 0.90 & 4.0 \\ %rms 0.00164
1997.93\,$^{(5)}$  &  8.4 & 0.43 & $0.6\times1.6$,\hspace{5.5pt}  33 & 0.38 & 0.5 \\ %rms 0.00017
1999.41\,$^{(5)}$  &  8.4 & 1.02 & $0.5\times1.5$,\hspace{4pt}  $-7$ & 0.92 & 0.5 \\ %rms 0.00016
1999.55\,$^{(2)}$  & 15.3 & 1.25 & $0.5\times0.8$,\hspace{9.5pt}   4 & 1.14 & 1.2 \\ %rms 0.00040
1999.89\,$^{a,(1)}$&  5.0 & 0.63 & $1.8\times2.4$,\hspace{0pt} $-60$ & 0.54 & 0.7 \\ %rms 0.00016
{\bf 2000.82}\,$^b$&  5.0 & 0.55 & $0.8\times0.9$,\hspace{0pt} $-28$ & 0.48 & 0.1 \\ %rms 0.00003
2001.17\,$^{(2)}$  & 15.3 & 0.65 & $0.4\times0.8$,\hspace{5.5pt}  31 & 0.54 & 0.9 \\ %rms 0.00033
\hline		
\end{tabular}
\flushleft
\footnotesize
Note: The array used was the VLBA, unless indicated by a footnote. Epochs in
bold face denote own data. a: Global array, b: VLBA+Eb\\
References: (1): \cite{1999bllp.conf..431B} \& in prep.\ (2005), (2): \cite{1998AJ....115.1295K} \&
\cite{2002AJ....124..662Z}, (3): \cite{2001ApJS..134..181J}, (4):
\cite{2000ApJS..128...17F}, (5): \cite{2001AaA...376.1090R}
\end{table}

\section{Observations and data analyses}\label{sec:obse}

\begin{figure*}[htbp]
\centering
\includegraphics[bb=0 0 2411 1890,angle=0,width=17cm,clip]
{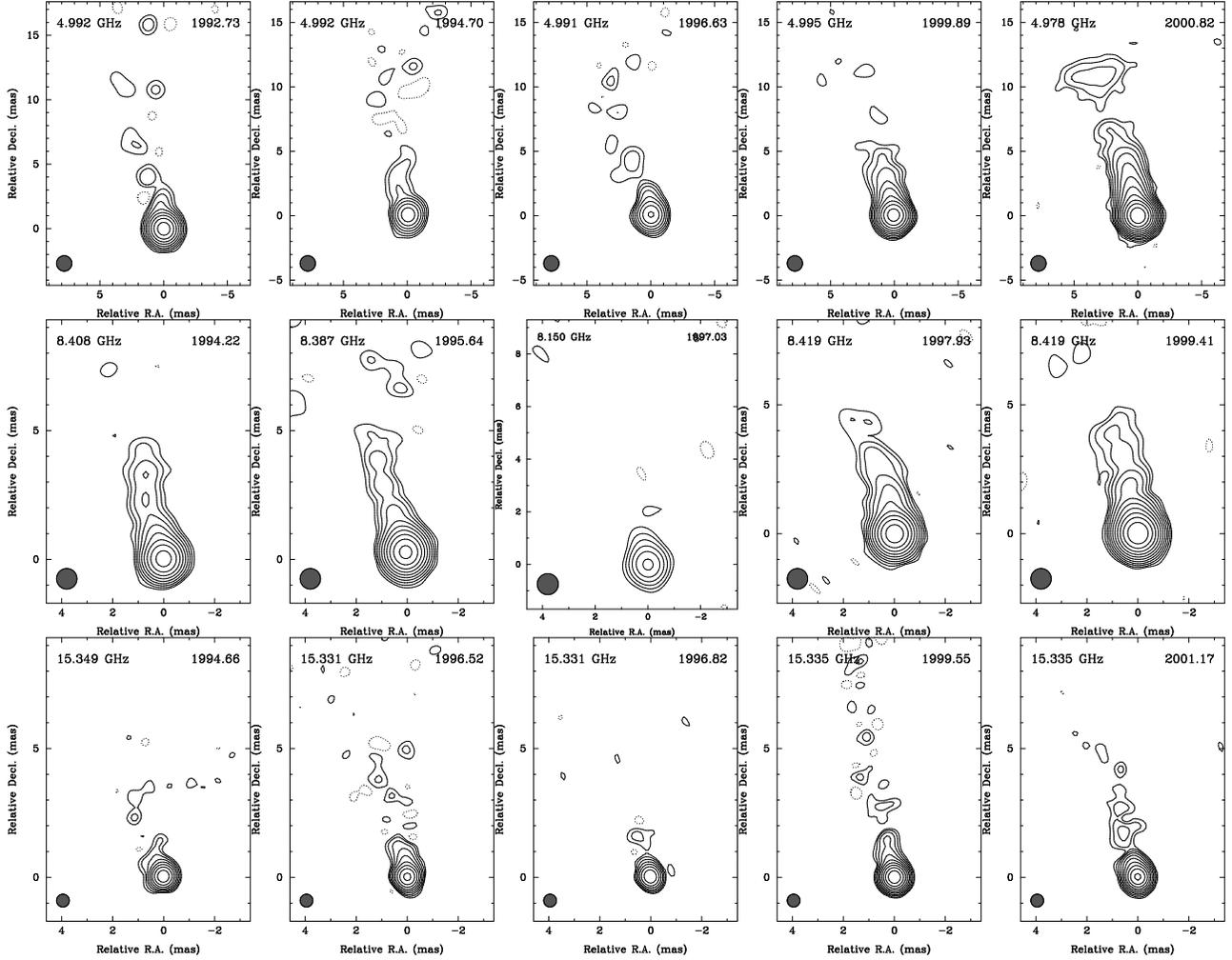} 
   \caption{All contour maps of 0716+714 at 5\,GHz (top row), 8\,GHz (middle) and 15\,GHz
   (bottom). The maps are convolved with circular beams
   of 1.2\,mas at 5\,GHz,  0.8\,mas at 8\,GHz and 0.5\,mas at
   15\,GHz. Total flux density, original beam size and the level of the lowest
   contour at three $\sigma$  are given in the observing log (Table~\ref{tab:obs}). Contours are
   increasing by steps of two.} 
   \label{fig:allmaps1}
\end{figure*}

\begin{figure*}[htbp]
\centering
\includegraphics[bb=0 0 1929 1890,angle=0,width=17cm,clip]
{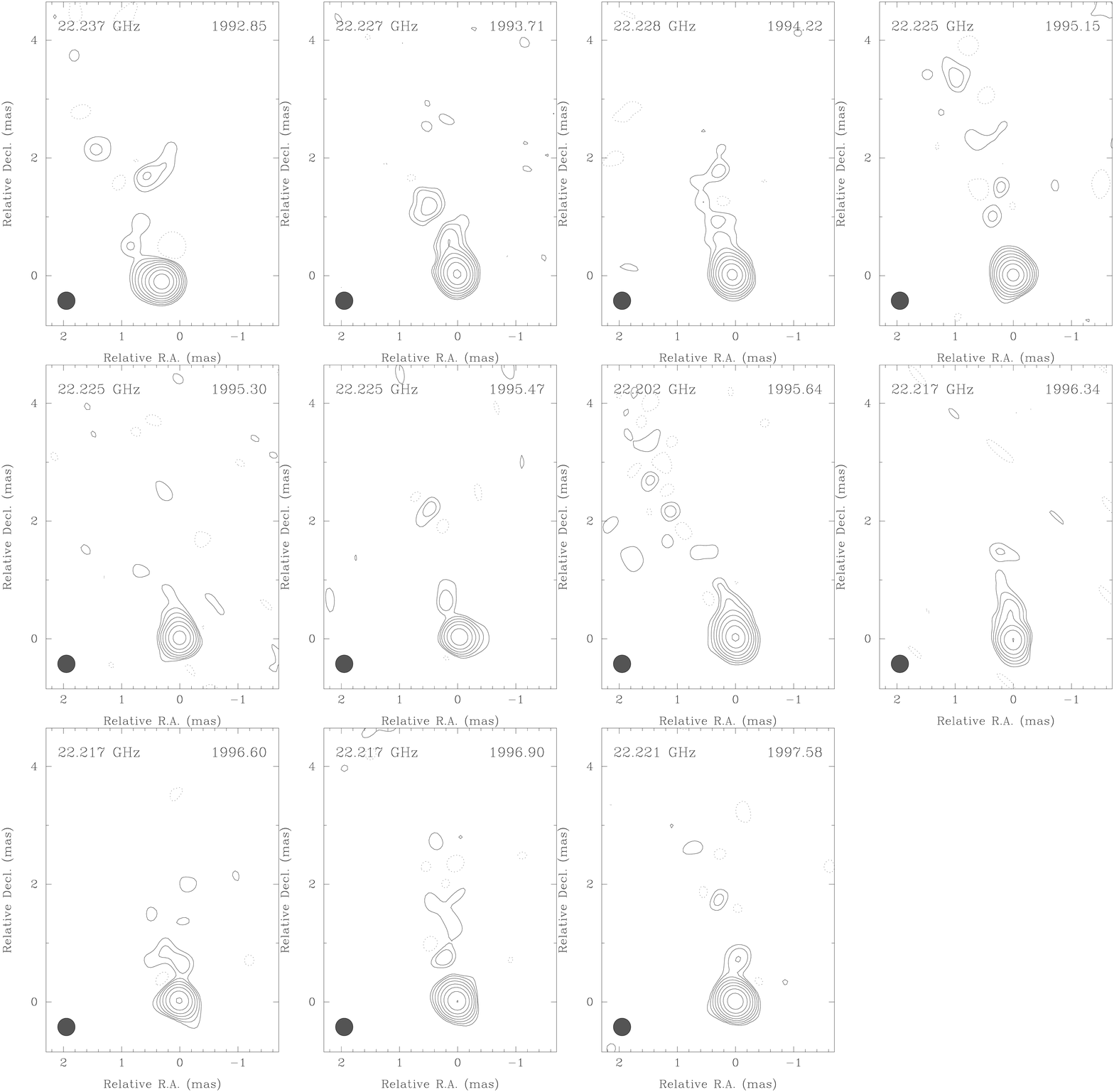} 
   \caption{All contour maps of 0716+714 at 22\,GHz. The maps are convolved with
   a circular beam of 0.3\,mas. Total flux density, original beam size and the
   level of the lowest contour at three $\sigma$ are given in the observing log
   (Table~\ref{tab:obs}). Contours are increasing by steps of two.}
   \label{fig:allmaps2}
\end{figure*}

All data were correlated in the standard manner using the MK\,III correlator in
Bonn and the VLBA correlator in Socorro. 
Part of the post-correlation analysis was done using NRAO's Astronomical Image
Processing System ({\sc Aips}) and the Caltech {\sc VLBI} Analysis Programs
(\citealt{1984ARA&A..22...97P}). Data from other observers were provided as
amplitude-calibrated and fringe fitted ($u,v$) FITS-files. The imaging of the source including phase and amplitude
self-calibration was done using the CLEAN algorithm (\citealt{1974A&AS...15..417H}) and
SELFCAL procedures in {\sc Difmap} (\citealt{1994BAAS...26..987S}). The self-calibration was done in steps of
several phase-calibrations followed by careful amplitude calibration. During the
iteration process the solution interval of the amplitude self-calibration was
shortened from intervals as long as the whole observational time down to minutes.
The resulting maps are presented in
 Figs.~\ref{fig:allmaps1}~\&~\ref{fig:allmaps2}. Here, the jet is clearly
visible to the north at P.A. $\sim 15\,^\circ$, and is slightly bent. At the
lower frequencies we can follow the jet up to a distance 
of 10\,mas to 15\,mas from the core, whereas at the higher frequencies the jet is
visible up to 3\,mas.

\subsection{Model fits}
\begin{figure}[htbp]
\centering
\includegraphics[bb=80 50 570 800,angle=-90,width=9.3cm,clip]
{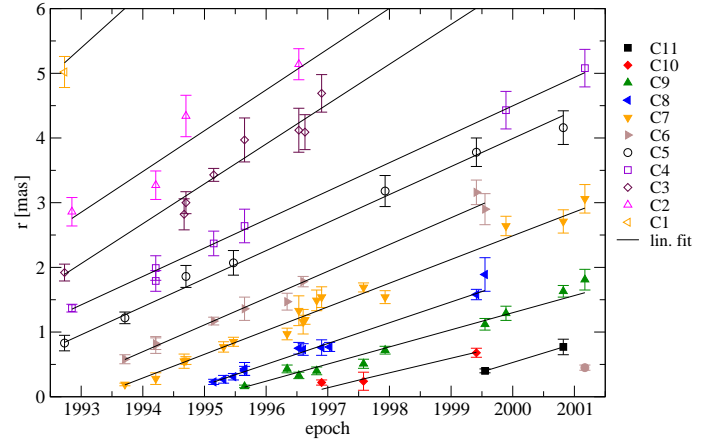} 
\caption{Core separation as a function of time for the individual modelfit
components. Data from all frequencies are combined. Possible frequency-dependent
position shifts are less than 0.1\,mas and are not corrected. The solid lines
show the linear fits to the path for each component. To show more clearly the
well-defined components we do not show two of the farthest data points of C1 at 
$r=11.2$\,mas and $r=11.9$\,mas and one data point of C2 at $r=6.9$\,mas.} 
\label{fig:r-time}
\end{figure}

After imaging in {\sc Difmap} we fitted circular Gaussian components
to the self-calibrated data to parameterise the source structure
seen in the VLBI maps. We used only circular components to reduce the number of
free parameters and to simplify the analysis. A 
summary of the Gaussian component parameters from this model fitting is given in
Table~\ref{tab:modelfits}.  For all model fits we used the brightest component
as a reference and fixed its position to (0,0). The positions of the other
components were measured relative to this component, which we assumed to be
stationary (see Sect.~\ref{sec:compid} for details). A maximum uncertainty of
15\,\% in the flux density was estimated from the uncertainties of the amplitude
calibration and from the formal errors of the model fits. The position error is
given by $\Delta r=\frac{\sigma\cdot \Theta}{2 S_{\rm P}}$
(\citealt{1989sira.conf..213F}), where $\sigma$ is the residual noise of the map
after the subtraction of the model, $\theta$ the full width at half maximum
(FWHM) of the component and $S_{\rm peak}$ the peak flux density. This formula
tends to underestimate the error, if the peak flux density is very high or the
width of the component is small. Therefore we included an additional error arising
from the position variations during the model fitting procedure. In
Fig.~\ref{fig:r-time}, the core distances of individual VLBI components
derived from our model fits at various frequencies are plotted against time.

\section{Results and discussion}\label{sec:discuss}

\subsection{Identification of the components}\label{sec:compid}
To investigate the kinematics in the jet of 0716+714, we cross-identified
individual model components along the jet using their distance from the VLBI
core, flux density and size. Since the observations were not phase-referenced,
the absolute position information was lost, and it is impossible to
tell which components, if any, are stationary. To register the images and to test
the stability of the VLBI-core position, we compared the core separations of the
jet components of each epoch with respect to the adjacent observations,
but we could not find any systematic position offsets. We also checked for
position shifts between the jet components at higher and lower 
frequencies due to opacity effects (e.g., \citealt{1998A&A...330...79L}). Inspection
of close or simultaneously observed frequency pairs does not show a systematic
frequency shift between higher and lower frequencies larger than 0.1\,mas
(position accuracy between 15\,GHz and 22\,GHz is $<0.1$\,mas). Therefore, we
conclude that the core position is 
stationary and that all component positions can be measured with respect to this
brightest component.

The cross identification of moving VLBI components between different times and
frequencies is not unambiguous and depends on the dynamic range of the
individual maps and on the time sampling. When we started our analysis we used
the following scenarios as working hypotheses:
\begin{enumerate}

\item Stationary core with stationary or slow-moving jet,

\item Stationary core with fast moving jet,

\item Non-stationary core with fast or slow-moving jet,

\item Non-monotonic motion of the jet.

\end{enumerate}

\begin{figure}[htbp]
\centering
\includegraphics[bb=49 44 686 508,angle=0,width=8cm,clip]
{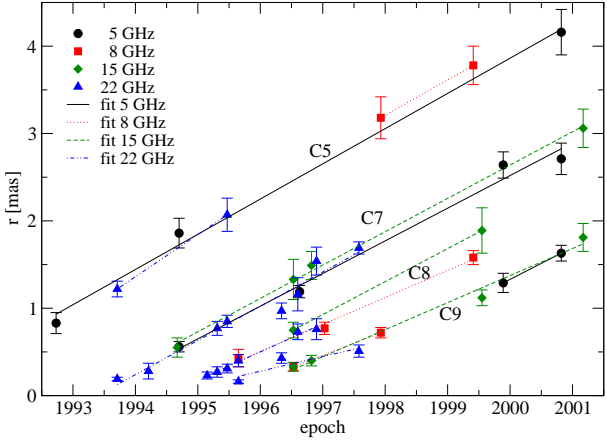} 
\caption{Core separation as a function of time for the components C5, C7, C8 and
C9 separated for the different frequencies. The solid lines
show the linear fits to the path for each component. Shifts between
the paths at different frequencies are visible, but are typically smaller than
the measured uncertainties.}
\label{fig:core-shift}
\end{figure}

As more data became available, we could rule out most of these identification
schemes, and we were left with a scheme that assumes a stationary core and
relatively fast component motion. The different schemes were
tested, first separately at each frequency and later with all modelfits
combined.

Supported by a graphical analysis, 
which is presented in Fig.~\ref{fig:r-time}, we could obtain a
satisfactory scenario for the kinematics in the jet of 0716+714. This
scenario consists of 11 identified superluminal components moving away from the
core. As an example of the procedure used for all components we show the
identification scheme for the separate frequencies for component C5, C7, C8 and
C9 in Fig.~\ref{fig:core-shift}. Again the frequency shifts between the
trajectories are typically smaller than 0.1\,mas to 0.2\,mas and are smaller
than the measurement uncertainty $\Delta r$.

Table~\ref{tab:beta_app} summarises the angular separation rate $\mu$ of the
VLBI components derived from the linear fits of $r(t)$ and the corresponding
apparent speed $\beta_{\rm app}=v_{\rm app}/c$ for an assumed redshift of
0.3.

\subsection{The kinematic model}
The components in this scenario move with 0.2\,mas\,yr$^{-1}$ to
0.6\,mas\,yr$^{-1}$ in the inner part of the VLBI jet ($r\leq3$\,mas) and
with up to 0.9\,mas\,yr$^{-1}$ in the outer regions. Due to the
inhomogeneous time sampling, it is sometimes difficult to identify
components over a long time interval. However, especially between 1993 and
1996, when we made many measurements spaced by only a few months, our
proposed component identification gives the simplest and most reasonable
fit to  the data. Since most of our data were obtained at high frequencies,
where the outer region of the jet is faint and already partly resolved
by the interferometer beam, the parameter of the corresponding jet components 
(at $r\geq4$\,mas) are not as well constrained as the inner jet components and
therefore they have larger positional errors. Despite this,
there is still a very distinct tendency that the older components (C1 -- C3), 
which now are located at large core separations, move faster than the components 
located in the inner jet (C4 -- C10). This behaviour is also visible in 
Table~\ref{tab:beta_app}, which shows a clear trend with systematically
decreasing speeds between component C1 (16\,$c$) and component C10 (4.5\,$c$). 
The coarse time sampling for the individual jet components unfortunately
does not allow us to fit acceleration to $r(t)$.
Future and more densely sampled VLBI observations are required to show whether the
inner jet components indeed move linearly.

\setcounter{table}{2}
\begin{table}[htbp]
\caption{Proper motions in the jet of 0716+714. The number of data points which were used for the fit for each
component is given in column 2. The last column gives the back-extrapolated
ejection dates of the components, which result from the linear fits.}
\label{tab:beta_app}
\begin{minipage}{\linewidth}
\renewcommand{\footnoterule}{}
\center
%\small
\begin{tabular}{lrcd{4}c}
\hline\hline
Id. & \# & $\mu$ [mas/yr] & \multicolumn{1}{c}{$\beta_{\rm app}\ (z=0.3)$} & Ejection date\\
\hline
 C1 &  4 & $0.86 \pm0.13$ & 16.13,2.36 & $1986.7 \pm1.3$ \\
 C2 &  5 & $0.63 \pm0.10$ & 11.87,1.82 & $1988.5 \pm1.0$ \\
 C3 &  8 & $0.62 \pm0.04$ & 11.59,0.76 & $1989.7 \pm0.3$ \\
 C4 &  7 & $0.44 \pm0.01$ &  8.27,0.24 & $1989.8 \pm0.1$ \\
 C5 &  7 & $0.43 \pm0.02$ &  8.16,0.29 & $1990.8 \pm0.2$ \\
 C6 &  9 & $0.41 \pm0.02$ &  7.80,0.33 & $1992.3 \pm0.1$ \\
 C7 & 17 & $0.37 \pm0.01$ &  6.89,0.19 & $1993.2 \pm0.1$ \\
 C8 & 11 & $0.32 \pm0.01$ &  6.04,0.20 & $1994.4 \pm0.1$ \\
 C9 & 10 & $0.26 \pm0.01$ &  4.98,0.27 & $1995.1 \pm0.1$ \\
C10 &  3 & $0.24 \pm0.01$ &  4.52,0.45 & $1996.5 \pm0.2$ \\
C11 &  2 & $0.29 \pm0.02$ &  5.48,0.55 & $1998.2 \pm0.3$ \\
\hline
\end{tabular}
\end{minipage}
\end{table}
\normalsize

\noindent 
The luminosity distance $d_{\rm L}$ was calculated adopting the following
relation

\begin{equation}
  d_{\rm L} =
\frac{c}{H_0}(1+z)\left[\eta(1,\Omega_{\rm m})-\eta(\frac{1}{1+z},\Omega_{\rm m})\right],
\end{equation}
where
\begin{eqnarray}
\eta(a,\Omega_{\rm m}) &=&
2\sqrt{s^3+1} \left[\frac{1}{a^4}-0.1540\frac{s}{a^3}+0.4304\frac{s^2}{a^2}\right.\nonumber\\
& & \left.+0.19097\frac{s^3}{a}+0.066941s^4\right]^{-\frac{1}{8}},\quad {\rm with}\nonumber\\
s^3&=&\frac{1-\Omega_{\rm m}}{\Omega_{\rm m}}\quad {\rm and}\quad a=\frac{1
}{1+z}, \nonumber
\end{eqnarray}

which is an analytical fit (\citealt{1999ApJS..120...49P}) to the luminosity
distance for flat cosmologies with a cosmological constant. For the range of
$0.2\leq \Omega_{\rm m}\leq 1.0$, the relative error of the distance is less than
0.4\,\% for any given redshift. Using

\begin{equation}
\beta_{\rm app} = \frac{\mu\ d_{\rm L}}{c\ (1+z)}
\end{equation}

for the transformation of apparent angular separation rates ($\mu$) into
spatial apparent speeds ($\beta_{\rm app}$)
(\citealt{1987slrs.work....1P}) and assuming a redshift of 0.3 for 0716+714, one
milliarcsecond corresponds to 4.4\,pc. The
measured angular separation rates then correspond to speeds
of 4.5\,$c$ to 16.1\,$c$.
For the components C3 to C9, for which the proper motion determination is the
most confident (over seven data points), the average motion is
$\mu=(0.41\pm0.11)$\,mas\,yr$^{-1}$ which corresponds to $\beta_{\rm
app}=7.7\pm2.1$. A preliminary analysis by \cite{2001IAUS..205...96T}, 
who used only part of the data presented here, gave very similar
speeds. A recently published kinematic analysis of the data from the VLBA
2\,cm Survey also measured motion of 10\,$c$ to 12\,$c$ in the jet of 0716+714
(\citealt{2004ApJ...609..539K}). The speeds found in our paper 
are lower than the 16\,$c$ to 21\,$c$ found by
\cite{2001ApJS..134..181J}\footnote{Jorstad et al. originally published
11\,$c$ to 15\,$c$ using $H_0=100$\,km\,s$^{-1}$\,Mpc$^{-1}$ and $q_0=0.1$. We corrected 
these numbers for the cosmological parameters used in this paper.}, who
observed the source over only a short time interval (3\,yr).
The difference between their and our results may be explained 
by slightly different (and unfortunately not unambiguous) 
map parameterisations using different Gaussian models or
by non-linear component motion, with possible 
acceleration for shorter periods of time. 

With an average apparent jet speed of about 8\,$c$ and likely
higher speeds of up to 15\,$c$ to 20\,$c$, 0716+714 displays considerably
faster motion than other BL\,Lac objects, for which speeds
of $\leq5\,c$ are regarded as normal (e.g., \citealt{1994ApJ...430..467V,
2000MNRAS.319.1109G, 2002evlb.conf..105R}). In this 
context it appears that 0716+714 is an extreme BL\,Lac object with
a jet speed (and Lorentz factor) much higher than regarded typical for BL\,Lac
objects and close to the speeds of 10\,$c$ to 20\,$c$ seen in quasars.

\subsection{Kinematics and geometry of the jet}\label{sec:kinematics}

Using the measured motion, $(11.6\pm0.8)\,c$, of C3, the fastest, best-constrained
jet component in our model, we can place limits on the jet speed and
orientation of 0716+174. Adopting 

\begin{equation}\label{eq:beta_app}
\beta_{\rm app}=\frac{\beta \sin\theta}{1-\beta \cos\theta}
\end{equation}

we find that the jet inclination that maximises 
the apparent speed, $\theta_{\rm max}$ is $4.9\,^\circ$. From that, one derives
a minimum Lorentz factor, $\gamma_{\rm min} = ( 1 + \beta_{\rm app}^2)^{1/2}$ of
$11.6$, which corresponds to a Doppler factor of  
$\delta_{\rm min}=\gamma_{\rm min}^{-1}(1 - \beta_{\rm min} \cos\theta_{\rm
max})^{-1} = 11.7$, where $\beta_{\rm min} = \sqrt{1-\gamma_{\rm
min}^{-2}}$. For smaller viewing angles ($\theta \rightarrow 0$) the Doppler
factor approaches $\delta = 2 \gamma$, which yields $\delta = 11.7$ to 23.4
at viewing angles between $4.9\,^\circ$ and $0\,^\circ$ for component C3.
Including the higher values found by \cite{2001ApJS..134..181J} and the highest speed
derived from our data (16.1\,$c$), a Doppler boosting factor of up to  
$\delta \geq 40$ appears possible. 

\begin{figure}[htbp]
\centering
\includegraphics[bb=14 14 740 525,angle=0,width=8.5cm,clip]
{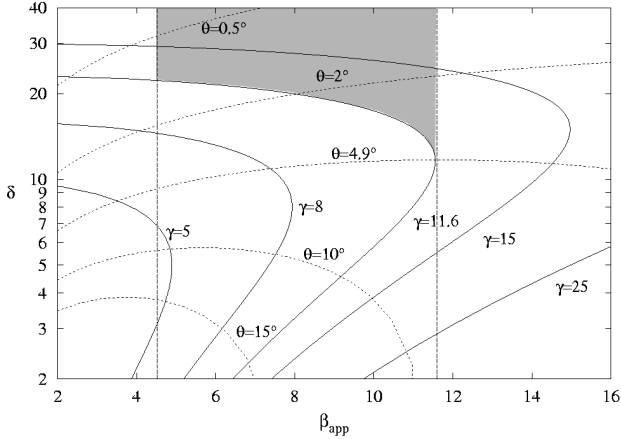}
   \caption{The Doppler factor versus the apparent
   speed for constant intrinsic Lorentz factor, $\gamma$, (solid
   lines) and constant viewing angles $\theta$ (dotted lines). The range
   of measured speeds is indicated by the dashed lines and the
   grey shaded area marks the possible values for $\gamma$ and
   $\theta$ (see text for details).}\label{fig:deltaplot}
\end{figure}

At cm-wavelengths, 0716+714 is a prominent intraday variable source, which shows
amplitude variations of 5\,\% to 20\,\% on time-scales 
of 0.25\,d to 2\,d (\citealt{1996AJ....111.2187W, 2003A&A...401..161K}).
Correlated radio-optical IDV observed in this source
(\citealt{1991ApJ...372L..71Q, 1996AJ....111.2187W, 1996ChA&A..20...15Q}) 
suggests that at least some fraction of the observed rapid variability has an
intrinsic source origin and cannot be attributed to refractive interstellar
scintillation (RISS) alone, as is done for some other IDV sources (e.g., 
\citealt{2001Ap&SS.278....5R, 2001A&A...367..770Q, 2001Ap&SS.278..113K}). The
presence of the observed broad-band correlations of the variability and recently
detected IDV at 9\,mm wavelength, where RISS should not play a dominant role due
to its $\lambda^{-2}$ dependence (cf. \citealt{2002PASA...19...14K,
2003A&A...401..161K}), further support a non negligible intrinsic contribution to the IDV 
in 0716+714. In the following we use an average typical brightness
temperature derived from the IDV observed at 6\,cm of $10^{15.5}$\,K to
$10^{17}$\,K. Slightly higher values of up to a few 
times $10^{18}$\,K were measured only occasionally. To bring this
brightness temperatures down to the inverse Compton limit of $10^{12}$\,K,
Doppler factors in the range of 
$\delta=(T_{\rm b}/10^{12}\,{\rm K})^\frac{1}{3}\approx 15$ to 50 are required.
Adopting these Doppler factors, we obtain jet Lorentz factors of $\gamma \approx
8$ to 25 and viewing angles of $\theta \leq 1\,^\circ$.

The relation between the Doppler factor, intrinsic Lorentz
factor, viewing angle and apparent speed is illustrated in
Fig.~\ref{fig:deltaplot}. The Doppler factors derived from IDV lie within the
grey shaded area.

If we now try to explain the observed 
range of apparent component speeds through variations of the viewing angle
(under the assumption of a constant Lorentz factor along the jet),
we cannot reach slow apparent speeds of $\sim 4.5$\,$c$
without violating the lower limit of the Doppler factor of 2.1 from
synchrotron self-Compton (SSC) models (\citealt{1993ApJ...407...65G}).
Therefore, we can exclude $\theta > 5\,^\circ$ and the only solution remaining
is to decrease the viewing angle which automatically leads to higher
Doppler factors. The grey shaded area in Fig.~\ref{fig:deltaplot} marks the
allowed region of Lorentz factors and viewing angles. It becomes obvious that we
need at least a Lorentz factor of $\gamma \approx15$ and a viewing angle of 
the VLBI jet of $\theta \approx2\,^\circ$ to explain the large
range of observed apparent speeds as an effect of spatial jet
bending. These values are consistent with those derived from IDV.

\subsection{Flux density evolution}
\begin{figure}[htbp]
\centering
\includegraphics[bb=10 44 686 515,angle=0,width=8.5cm,clip]
{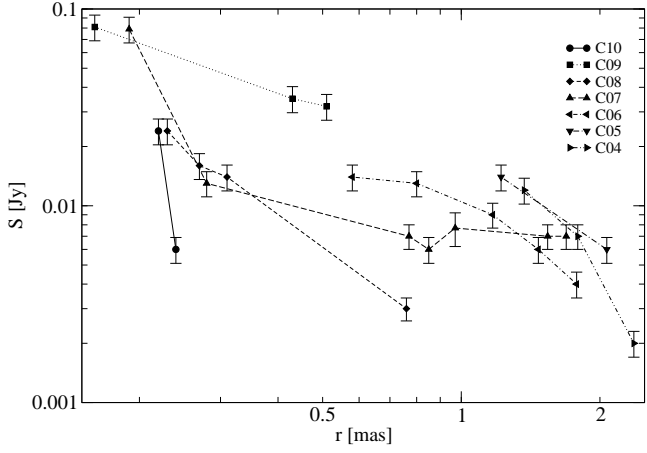}
   \caption{Flux density of the VLBI components at 22\,GHz plotted versus
   core separation. The components fade as they travel down the jet.}  \label{fig:flux-r}
\end{figure}

\begin{figure}[htbp]
\centering
\includegraphics[bb=21 44 686 515,angle=0,width=8cm,clip]
{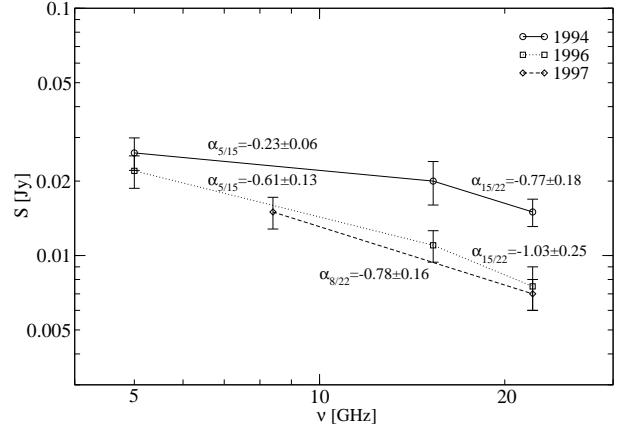}
\caption{Spectral evolution of component C7.}  
\label{fig:spectrum}
\end{figure}

In Fig.~\ref{fig:flux-r} we show the evolution of the flux density of
VLBI components at 22\,GHz with their increasing separation from the core. We
show this frequency because a large fraction of the data were obtained at
22\,GHz (11 epochs). With the exception of a few data points, the components
fade as they travel down the jet. This agrees qualitatively with the theory of a
conically expanding jet (e.g., \citealt{1979ApJ...232...34B}). As the VLBI
components expand they become optically thin and their spectra steepen.
This is illustrated for component C7 in Fig.~\ref{fig:spectrum}. 
The spectral index $\alpha$ is defined as $S \propto \nu^\alpha$. Knowing the
turnover frequency and the size of a jet component allows us to calculate the strength of
the magnetic field in the jet (e.g., \citealt{1983ApJ...264..296M}) as
\begin{equation}\label{eq:bfield}
B = 10^{-5}\ b(\alpha)\ \frac{ \theta^4 \nu_{\rm m}^5 \delta}{S_{\rm m}^2\,(1+z)}\
{\rm gauss},
\end{equation}
where $b(\alpha)$ is a tabulated parameter dependent on the spectral index (Tab.~1 in
\citealt{1983ApJ...264..296M}) with a value ranging from 1.8 to 3.8 for
optically thin emission ($\alpha=-0.25$ to $-1.0$), $S_{\rm m}$ (in Jy) is 
the flux density at the turnover frequency $\nu_{\rm m}$ (in GHz) and $\theta$
(in mas) is the size of the component. If we calculate B for each of our modelfit
components, without correcting for the Doppler factor $\delta$, we get the
apparent magnetic field along the jet. Since we do not have adequate spectra for
all positions along the jet we will consider a decrease of the turnover
frequency along the jet proportional to $r^p$, to correct for the expansion of
the component. The best results are obtained when we start with a turnover
frequency of 4\,GHz for the inner jet ($r\approx0.2$\,mas), that is plausible
from the early spectrum (1994) of component C7 (Fig.~\ref{fig:spectrum})
though the turnover is not actually observed, and use  $p=-1.3$, which is in
good agreement with the relativistic jet model (\citealt{1985ApJ...298..114M}).
For all components along the jet we derive magnetic fields in the range of
$10^{-4}\,\delta$\,G to $10^{-7}\,\delta$\,G. Since the minimum magnetic field strength that should be present due the energy
density of the cosmic microwave background is a few microGauss (e.g.,
\citealt{1969A&A.....3..468V}) the $10^{-7}\,\delta$\,G  already implies a lower
limit of the Doppler factor of about 10. At this stage the core component is excluded from the
analysis because both the source size and the turnover frequency are poorly
known.

Assuming equipartition between the energy of the particles, $E_{\rm e}$, and the
energy of the magnetic field, $E_{\rm B}$, one can also derive the minimum
magnetic field from the synchrotron luminosity, $L$.
\begin{eqnarray}
L & =& 4\pi\ d_{\rm L}^2\ (1+z)\int_{\nu_1}^{\nu_2}\ S\ d\nu \\
E_{\rm e} & = & f(\alpha,\nu_1,\nu_2)\ L\ B^{-1.5},
\end{eqnarray}
where $f(\alpha,\nu_1,\nu_2)$ is a tabulated function (e.g.\ \citealt{1970ranp.book.....P})
and $\nu_1$ and $\nu_2$ are the upper and lower 
cutoff frequencies of the synchrotron spectrum with typically values of
$10^7$\,Hz to $10^{11}$\,Hz. Taking the energy density of the magnetic field
 to be
$1/8\pi\,B^2$ and assuming spherical symmetry the total energy of the source is
\begin{eqnarray}\label{eq:etot}
E_{\rm tot} & = & (1+k) E_{\rm e} + E_{\rm B}\\\nonumber
 & = & (1+k)\,f(\alpha,\nu_1,\nu_2)\ L\ B^{-1.5} + \frac{1}{6}\,R^3\,B^2,
\end{eqnarray}
where $k$ is the energy ratio between the heavy particles and the electrons and
$R$ is the size of the component. The ratio $k$ depends on the
mechanism of generation of the relativistic electrons, which is unknown at the
present. It can range from $k\approx1$ for an electron-positron plasma up to
$k\approx2000$, if most of the energy is carried in the protons. We used 
$k\approx 100$ in our calculations, which seems to be a reasonable value (e.g.,
\citealt{1970ranp.book.....P}).
Given that $E_{\rm e}\propto B^{-1.5}$ and $E_{\rm B}\propto B^2$, $E_{\rm
tot}$ has a minimum and we find 
\begin{eqnarray}\label{eq:bmin}
B_{\rm min} & = & \left(\,\frac{9}{2}\,(1+k)\,f(\alpha,\nu_1,\nu_2)\ L\
R^{-3}\,\right)^{2/7}\\\nonumber
 & = & 5.37\cdot 10^{12}\ (S_{\rm m}\ \nu_{\rm m}\ d_{\rm L}^2\ R^{-3})^{2/7}.
\end{eqnarray}

with $k=100$, $f(-0.5,10^7,10^{11})=1.6\cdot10^7$ and $S_{\rm m}$ in Jy,
$\nu_{\rm m}$ in GHz and $R$ in cm. Using $z=0.3$ and the same values for  
$S_{\rm m}$, $\nu_{\rm m}$ and $\theta$ as in Equation~\ref{eq:bfield} we
obtain magnetic fields between $10^{-4}$\,G and $10^{-2}$\,G for the jet. These
values are larger than those derived before, which means that either the
particle energy dominates and we do not have equipartition or that the emission
is relativistically beamed. In the latter case, the different dependence of the
magnetic field on the Doppler factor in Equation~\ref{eq:bfield} ($B \propto
\delta$) and in Equation~\ref{eq:bmin} ($B_{\rm min}\propto
\delta^{2/7\alpha+1}$) can be used to calculate the Doppler factor from 
$B_{\rm min}/B=\delta^{2/7\alpha}$. Comparing the results from all jet
components yields Doppler factors of 4 to 15 with a mean value of 9. These
values are slightly lower than those derived from the kinematics, but given that
the calculations are based on many assumptions the values agree reasonably. 

\begin{figure}[htbp]
\centering
\includegraphics[bb=14 14 491 725,angle=0,width=8cm,clip]
{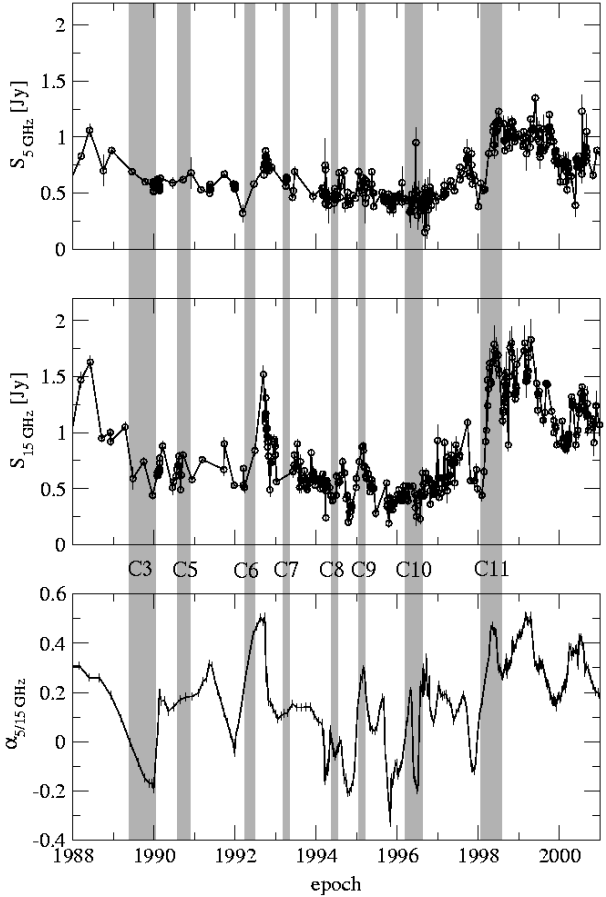}
   \caption{The long term flux density variability of 0716+714 as measured at
   Effelsberg and with the Michigan radio telescope at 5\,GHz (top) and 15\,GHz
   (middle) and the spectral index (bottom), $\alpha_{5/15\,{\rm GHz}}$. The
   shaded areas mark the ejection dates of the VLBI components and their
   uncertainties.}
\label{fig:lightcurve}
\end{figure}

In Fig.~\ref{fig:lightcurve} we show the long-term radio variability and the
spectral index of 0716+714 over the time range in which the VLBI components were
born. We show a combined data set with flux density measurements
at 5\,GHz and 15\,GHz from the UMRAO flux density monitoring
program (\citealt{2003AAS...202.1801A}) and from our flux-density monitoring
performed with the 100\,m radio telescope at Effelsberg
(\citealt{2000A&AS..145....1P}). A detailed discussion 
of the flux density variability using these and other data was recently done by
\cite{2003A&A...402..151R}.
Here, we restrict the discussion to a possible correlation of the radio
variability with the ejection of new jet components. The
ejection dates of new VLBI components are indicated by the grey
shadowed areas. Their widths correspond to the
uncertainty in the back-extrapolated ejection dates (see Table~\ref{tab:beta_app}). 
Although these uncertainties are rather large and the time sampling of
the light curves is not always dense enough, a weak correlation between the
ejection of new components and the flares in the radio bands 
is obvious. Each of the shaded areas either lies at or shortly before the time
of a flux density increase of at least one of the two observing bands.

The two outbursts in mid 1992 and early 1995, which are best seen in
the 2\,cm band, are both surrounded by two new components,
but the time sampling is too sparse to draw any further conclusions.
In the lower panel of Fig.~\ref{fig:lightcurve} we calculated the spectral
index between 5\,GHz and 15\,GHz from the two light curves. Since the light
curves were not measured simultaneously we searched for the closest pairs in the
data set and calculated the spectral index from these pairs. The mean separation
between two points is about eight days. The graph itself represents an
eight-point running average to reduce the noise in the data. This figure shows
an obvious correlation between the spectral index variations and the ejection of
a new VLBI component. Seven out of eight ejections (except C9) of new VLBI components are
accompanied by a flattening of the source spectrum. The flattening is in good
agreement with an expanding component that becomes optically thin first at the
higher frequency and later at the lower frequency (e.g.,
\citealt{1985ApJ...298..114M}).

We note that at the times of particularly dense
time sampling (after 1994) a number of minor radio flares are visible, which are
not related to the ejection of any of the known VLBI components. According to
the light-house model and other related helical jet models (e.g., 
\citealt{1992A&A...255...59C}; \citealt{1994A&A...290..357R}) it is possible
that initial outbursts, which are related to the ejection of new jet components
are followed by secondary flux density variations, which are not or only are
indirectly related to the component ejection. Of course, it is also possible
that we missed some jet components in our infrequently sampled VLBI monitoring.

To test the significance of the correlations found by eye, we placed eight
fields with a similar width randomly over the light curves between 1989 and 1999
and determined how many are coincident with a flare in the light curve or a
flattening of the spectrum. For the light curves this test yielded a 50\,\%
probability of our results occurring at random. The probability of
measuring a flattening of the spectrum in seven out of eight components selected
at random from our spectrum is 5.7\,\%. Generally a result is excepted to be
statically significant if the probability is below 5\,\%, which means that the
correlations of the light curve with the ejection
dates are likely to be a coincidence, but given that we might have missed a
component the correlation of the spectral index is more significant. A coordinated and much denser sampled
multi-frequency flux density and 
VLBI monitoring is necessary to study in more detail such outburst-ejection
relations.  

\subsection{Connection with $\gamma$-ray flares}

0716+714 was detected at  $\gamma$-rays by the EGRET detector on board the
{\it Compton Gamma Ray Observatory}
(\citealt{1999ApJS..123...79H,1997ApJ...481...95M}). Since it is proposed that
the $\gamma$-ray emission, like the radio emission, is Doppler boosted
(\citealt{1994ApJS...90..945D}), the $\gamma$-ray detection of 0716+714 yields
further evidence for a high Lorentz factor in the jet. 

Recent studies about the connection between $\gamma$-ray sources and superluminal
VLBI-components suggest that these sources have higher average jet speeds than
sources which are not detected at $\gamma$-rays
(\citealt{2001ApJ...556..738J,2004ApJ...609..539K}). Moreover,
\cite{2001ApJ...556..738J} found a correlation between the ejection of new VLBI
components and the appearance of  $\gamma$-ray flares, suggesting that the radio
and the $\gamma$-ray emission originate within the same shocked area in the
relativistic jet. The authors report a $\gamma$-ray flare of 0716+714 in 1992.2
which appears to be coincident with a peak in the linear polarization light-curve
at 15\,GHz. Unfortunately, their 0716+714 VLBI data does not cover this period, but
with our larger sample we found a new VLBI component (C6) with an extrapolated
ejection date of $1992.3\pm0.1$ which is in good agreement with the reported
$\gamma$-ray flare. Although we register three additional ejections up to 1996 and
0716+714 was observed several times by EGRET in this period, no more flares were
detected. But given that the $\gamma$-ray measurements have large uncertainties and
only 18 data points were recorded during these four years, this seems not
exceptional. Similar outburst ejection events were also found in e.g, 0528+134
(\citealt{1995PNAS...9211377K}) and 0836+710 (\citealt{1998A&A...334..489O}).
Assuming that the correlation between the new component, C6, and the 
1992.2 flare is real, this supports the idea that the $\gamma$-ray emission
originates from the same region from which we observe the radio emission (e.g.,
\cite{1994ApJ...421..153S}).

\subsection{Variation of the ejection position angle and core flux density}
\begin{figure}[htbp]
\centering
\includegraphics[bb=29 45 775 515,angle=0,width=8.5cm,clip]
{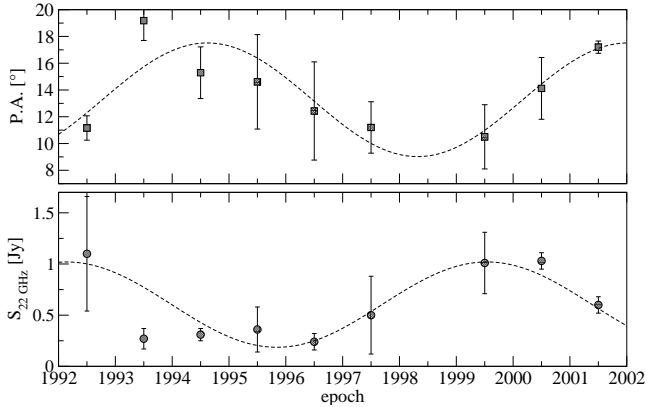}
   \caption{Top: Position angle of the central 1\,mas of the VLBI jet
    grouped in one year time bins. The dashed line is a 
    sinusoidal fit to the data. The position angle of the
    inner jet (ejection angle) varies with a period of $\sim7.4$ years. Bottom:
    One year bins of the 22\,GHz core flux density. The dashed line represents a
    sinusoidal fit with a fixed period of 7.4 years.} 
\label{fig:ejection_angle}
\end{figure}

The visual inspection of position angle of the innermost
portion of the VLBI jet in the maps of Figs.~\ref{fig:allmaps1}~\&~\ref{fig:allmaps2}
indicates a variation of the P.A. of the jet near the core with time. In order
to quantify this, we used the Gaussian models and fitted a straight line to the
first milliarcsecond of the jet. For each observing epoch, the position angle 
of this line provides a good estimate of the direction of the inner portion
of the VLBI jet. In Fig.~\ref{fig:ejection_angle} (top panel) we plot the P.A.
grouped in one year time bins.

A sinusoidal fit to the data reveals a period of
($7.4\pm1.5$)\,yr. The fit yields a reduced $\chi_{\rm r}^2$ of 1.55, which is
much better than the $\chi_{\rm r}^2$ of 4.28, obtained for a linear fit.
Unfortunately our data cover only one period of the variability and 
it will be interesting to see if the trend holds in the future. A possible
explanation for such periodic variation would be a precessing footpoint of the
jet. This should also result in a flux density variation of the VLBI core, and
indeed, there is a tendency that the core flux density at 22\,GHz is
consistent with a periodic variation of $\sim 7$\,yr (see
Fig.~\ref{fig:ejection_angle}, bottom panel).
%The flux density measurements at
%the lower frequencies (5\,GHz, 8\,GHz, and 15\,GHz) are scaled to 22\,GHz using
%the spectral indices presented in Fig.~\ref{fig:lightcurve}. We note that
%these spectral indices are derived from the whole source, but since the total flux
%density of 0716+714 is dominated by the VLBI-core ($\sim 80$\,\%), the
%uncertainties arising from that are negligible.
Recently \cite{2003A&A...402..151R} published an analysis of the optical and radio flux
density variability of 0716+714 and found a periodicity of 5.5\,yr to 6.0\,yr
for the flux-density variations in the radio regime. This is compatible within
the error bars with the periodicity of the P.A. in the inner region of the jet. 
Based on the geometrical considerations in Sect.~\ref{sec:kinematics}, the
apparent peak to peak variation of $\approx 7\,^\circ$ of the ejection angle
corresponds to a small change of only $7\,^\circ \sin\theta \approx 0.6\,^\circ$
of the jet direction.

Such small variations of a few degrees in the rest frame of the source in
several years have recently been detected also in other AGN, e.g.
3C\,345 (\citealt{1986ApJ...308...93B}), 3C\,279
(\citealt{1993A&A...279...83C}), 3C\,273 (\citealt{2001pfrg.conf..184K}) and
BL\,Lac (\citealt{2003MNRAS.341..405S}). The most common explanation is a
precessing jet, which could be caused by a binary super-massive black hole
system (\citealt{1986ApJ...308...93B,1992A&A...257..489H}) or a 
warped accretion disc (\citealt{2003ApJ...591L.119L} and ref. therein). However,
hydrodynamic effects also can lead to a precession of the jet (e.g.,
\citealt{2002ApJ...572..713H}). In this case the precession is introduced by a
potentially strong oblique internal shock that arises from asymmetric
perturbation of the flow.

\subsection{Redshift dependence of the results}
Since all the derived parameters of the kinematics and geometry scale with
the distance to 0716+714, knowing the redshift is of great importance. The
distance to 0716+714 and therewith the speeds of the jet components scale nearly
linearly with $z$ at these low redshifts. Thus a redshift decrease by a factor
of three would also decrease the Doppler factor by three. The dependence of
brightness temperature derived from IDV is $T_{\rm b} \propto 
\left(\frac{z}{1+z}\right)^2$ (e.g., a decrease from  $z=0.3$ to $z=0.1$ would
reduce $T_{\rm b}$ by 6.4). Since, $T_{\rm b}\propto \delta^3$ this would
increase the discrepancy between the Doppler factor, which is needed to reduce
$T_{\rm b}$ and the Doppler factor derived from the kinematics. An increase in
redshift would bring the derived parameters closer together.

Our assumption of the redshift of  $z\geq0.3$ is based on the non-detection of
an underlying host galaxy by \cite{1996AJ....111.2187W}. BL\,Lac host galaxies
have typical absolute  magnitudes of $M_{\rm R}\approx -24$ and an effective
radius of  $r_{\rm c}\approx10$\,kpc (\citealt{1997ApJ...491..146J} and ref.
therein). The lower limit of Wagner et al. corresponds to an absolute magnitude
of $M_{\rm R}= -20.4$, which would be an unusually faint BL\,Lac host and so is
a very conservative lower limit. An upper limit for the redshift can be given
from the lack of absorption by foreground galaxies, which results in a redshift
of $z\leq0.5$ (Wagner priv.\ comm.). At present it is unclear if the
tentative X-ray detection of a spectral line near 5.8\,keV
(\citealt{2004A&A.....0....0K}) can be identified with the Fe\,K$_\alpha$ line.
If so, the redshift of the line would indicate a distance of $z=0.10\pm0.04$,
much closer than expected from the optical measurements. Possible ways out of
this difficulty could invoke a blue shift of the line or
some external emission process not related to the nucleus of 0716+71. A solid
confirmation of the X-ray line in future observations is urgently needed.

\section{Conclusions}\label{sec:conclusion}
We analysed 26 epochs of VLBI data at 4.9\,GHz, 8.4\,GHz, 15.3\,GHz,
and 22.2\,GHz observed over 10\,yr, between 1992 and 2001, and derived a new
kinematic scenario for the jet in 0716+714. In this scenario the components move
 with apparent speeds of 4.5\,$c$ to 16.1\,$c$ ($z=0.3$). These speeds are
atypically fast for BL\,Lac objects, which typically exhibit speeds of
$\leq5\,c$ (e.g., \citealt{1994ApJ...430..467V, 2000MNRAS.319.1109G,
2002evlb.conf..105R}). Since this analysis is based on much denser sampled VLBI
data than previous studies, we are convinced
that we can rule out the somewhat slower scenarios which were presented
previously (e.g., \citealt{1988A&A...206..245W, 1992vob..conf..225S,
1998A&A...333..445G}). We find a periodic variation of the P.A. of
the innermost jet which could be a sign of jet precession on a time-scale
of ($7.4 \pm 1.5$)\,yr. If this can be confirmed in future observations the
higher speeds found by \cite{2001ApJS..134..181J} might be explained by
non-linear motion of the jet components.

No correlation was found between the component ejection and radio flux density
flares at the cm-wavelengths. There seems to be a weak correlation between the
ejection of new components and the flattening  of the radio spectrum. For one of
the new components (C6) we could find a reported $\gamma$-ray flare and if the
ejection and the flare are really connected this would support the idea that the
$\gamma$-ray emission originates from the same region from which we observe the
radio emission. 

From the component motion in the jet, we obtain a lower limit for the Lorentz
factor of 11.6 and a maximum angle to the line of sight of $4.9\,^\circ$. To
explain the large range of observed apparent speeds as an effect of spatial jet 
bending, a Lorentz factor of $\gamma>15$ and a viewing angle of
the VLBI jet of $\theta <2\,^\circ$ are more likely. Under these 
circumstances, the Doppler factor would be $\delta\approx 20$ to 30. Such high
Doppler factors are indeed required to explain the high
apparent brightness temperatures of up to $10^{17}$\,K derived
from intraday variability at cm-wavelengths.

\begin{acknowledgements}
We thank S. Jorstad and A. Marscher for providing their VLBA data for
reanalysis. We also thank the group of the VLBA 2\,cm Survey and the group of
the CJF Survey for providing their data. We appreciate the use of the flux
density monitoring data from the UMRAO data base. We thank the anonymous referee
for helpful comments and suggestions. This work made use of the
VLBA, which is an instrument of the National Radio Astronomy Observatory, a
facility of the National Science Foundation, operated under cooperative
agreement by Associated Universities, Inc., the European VLBI Network, which
is a joint facility of European, Chinese, South African and other radio
astronomy institutes funded by their national research councils. This work is
also based on observations with the 100\,m radio telescope of the MPIfR
(Max-Planck-Institut f\"ur Radioastronomie) at Effelsberg. We gratefully
acknowledge the VSOP Project, which is led by the Japanese Institute of Space
and Astronautical Science in cooperation with many organisations and radio
telescopes around the world. 
\end{acknowledgements}

\bibliographystyle{aa} % style aa.bst
%\bibliography{references}

\setcounter{table}{1}
\begin{table}[htbp]
\caption{Results from Gaussian Model fitting and component parameters. $S_{\rm
peak}$ is the peak flux density, $r$ and $\phi$ are the distance and the P.A.
measured from the core and $\theta$ is the FWHM of the Gaussian component.}
\label{tab:modelfits}
\begin{minipage}{\linewidth}
\renewcommand{\footnoterule}{}
\centering
\scriptsize
\begin{tabular}{l.d{2}d{2}d{2}d{2}l}
\hline
\multicolumn{1}{c}{Epoch} &
\multicolumn{1}{c}{$\nu$} &
\multicolumn{1}{c}{$S_{\rm peak}$} & 
\multicolumn{1}{c}{$r$} &
\multicolumn{1}{c}{$\phi$} &
\multicolumn{1}{c}{$\theta$} &
\multicolumn{1}{c}{Id.$^a$}\vspace{1pt}\\
 &
\multicolumn{1}{c}{[GHz]} &
\multicolumn{1}{c}{$[\frac{mJy}{beam}]$} & 
\multicolumn{1}{c}{[mas]} &
\multicolumn{1}{c}{[$^\circ$]} &
\multicolumn{1}{c}{[mas]} & \vspace{1pt}\\
\hline
1992.73 &  5.0 & 596,89.4 & \multicolumn{1}{c}{\hspace{3mm}--} & \multicolumn{1}{c}{--} & 0.10,0.02 & Core\\ 
& &  37, 5.5 &  0.83,0.12 & 17, 8 & 0.14,0.03 & C5\\ 
& &  19, 2.8 &  1.92,0.13 &  8, 3 & 0.60,0.12 & C3\\ 
& &   8, 1.2 &  5.02,0.24 & 17, 2 & 0.93,0.19 & C1\\ 
& &   7, 1.0 & 11.46,0.49 & 13, 2 & 3.35,0.67 & X\\ 
1992.85 & 22.2 & 708,106.2 &  \multicolumn{1}{c}{\hspace{3mm}--} & \multicolumn{1}{c}{--} & 0.18,0.04 & Core\\ 
& &  12, 1.8 &  1.37,0.06 & 12, 2 & 0.10,0.02 & C4\\ 
& &  16, 2.4 &  2.86,0.22 &  0, 4 & 1.48,0.30 & C2\\ 
1993.71 & 22.2 & 272,40.8 &  \multicolumn{1}{c}{\hspace{3mm}--} & \multicolumn{1}{c}{--} & 0.05,0.01 & Core\\ 
& &  79,11.8 &  0.19,0.02 & 15, 5 & 0.05,0.01 & C7\\ 
& &  14, 2.1 &  0.58,0.07 & 14, 6 & 0.12,0.02 & C6\\ 
& &  14, 2.1 &  1.22,0.09 & 20, 4 & 0.23,0.05 & C5\\ 
1994.21 & 22.2 & 309,46.3 &  \multicolumn{1}{c}{\hspace{3mm}--} & \multicolumn{1}{c}{--} & 0.09,0.02 & Core\\ 
& &  15, 1.9 &  0.28,0.09 & 28,18 & 0.01,0.00 & C7\\ 
& &  13, 1.9 &  0.80,0.13 & 15, 9 & 0.42,0.08 & C6\\ 
& &   7, 1.0 &  1.79,0.16 & 14, 5 & 0.38,0.08 & C4\\ 
1994.21 &  8.4 & 266,39.9 & \multicolumn{1}{c}{\hspace{3mm}--} & \multicolumn{1}{c}{--} & 0.14,0.03 & Core\\ 
& &  25, 3.7 &  0.82,0.09 & 13, 6 & 0.42,0.08 & C6\\ 
& &  10, 1.5 &  1.99,0.19 & 17, 5 & 0.76,0.15 & C4\\ 
& &   9, 1.3 &  3.27,0.22 & 12, 3 & 0.84,0.17 & C2\\ 
& &   4, 0.6 &  4.70,0.43 & 14, 5 & 1.48,0.30 & X\\ 
1994.67 & 15.3 & 413,61.9 &  \multicolumn{1}{c}{\hspace{3mm}--} & \multicolumn{1}{c}{--} & 0.07,0.01 & Core\\ 
& &  20, 4.0 &  0.55,0.11 & 14,11 & 0.64,0.13 & C7\\ 
& &  14, 2.1 &  2.82,0.24 & 14, 5 & 1.68,0.34 & C3\\ 
1994.70 &  5.0 & 284,42.6 &  \multicolumn{1}{c}{\hspace{3mm}--} & \multicolumn{1}{c}{--} & 0.18,0.04 & Core\\ 
& &  26, 3.9 &  0.56,0.06 &  7, 6 & 0.21,0.04 & C7\\ 
& &  13, 1.9 &  1.86,0.17 & 18, 5 & 0.77,0.15 & C5\\ 
& &  12, 1.8 &  3.00,0.17 & 15, 3 & 0.72,0.14 & C3\\ 
& &   7, 1.0 &  4.34,0.32 &  8, 4 & 1.45,0.29 & C2\\ 
1995.15 & 22.2 & 724,108.6 &  \multicolumn{1}{c}{\hspace{3mm}--} & \multicolumn{1}{c}{--} & 0.07,0.01 & Core\\ 
& &  24, 3.6 &  0.23,0.04 & 12, 9 & 0.07,0.01 & C8\\ 
& &   9, 1.3 &  1.17,0.06 & 12, 3 & 0.07,0.01 & C6\\ 
& &   2, 0.3 &  2.37,0.19 & 14, 4 & 0.15,0.03 & C4\\ 
& &  10, 1.5 &  3.43,0.10 & 15, 1 & 0.20,0.04 & C3\\ 
1995.31 & 22.2 & 421,63.1 &  \multicolumn{1}{c}{\hspace{3mm}--} & \multicolumn{1}{c}{--} & 0.10,0.02 & Core\\ 
& &  16, 2.4 &  0.27,0.06 & 12,11 & 0.10,0.02 & C8\\ 
& &   7, 1.0 &  0.77,0.08 & 14, 6 & 0.10,0.02 & C7\\ 
1995.47 & 22.2 & 174,26.1 &  \multicolumn{1}{c}{\hspace{3mm}--} & \multicolumn{1}{c}{--} & 0.04,0.01 & Core\\ 
& &  14, 2.1 &  0.31,0.05 & 29, 9 & 0.07,0.01 & C8\\ 
& &   6, 0.9 &  0.85,0.07 & 18, 4 & 0.06,0.01 & C7\\ 
& &   6, 0.9 &  2.07,0.19 & 23, 5 & 0.43,0.09 & C5\\ 
1995.65 & 22.2 & 220,33.0 &  \multicolumn{1}{c}{\hspace{3mm}--} & \multicolumn{1}{c}{--} & 0.07,0.01 & Core\\ 
& &  81,12.1 &  0.16,0.02 & 10, 6 & 0.05,0.01 & C9\\ 
& &  29, 4.3 &  0.40,0.07 & 13,10 & 0.32,0.06 & C8\\ 
1995.65 &  8.4 & 252,37.8 &  \multicolumn{1}{c}{\hspace{3mm}--} & \multicolumn{1}{c}{--} & 0.11,0.02 & Core\\ 
& &  32, 4.8 &  0.43,0.10 &  9,12 & 0.10,0.02 & C8\\ 
& &   8, 1.2 &  1.36,0.18 & 15, 7 & 0.53,0.11 & C6\\ 
& &   3, 0.4 &  2.64,0.26 & 20, 5 & 0.40,0.08 & C4\\ 
& &   5, 0.8 &  3.97,0.34 & 15, 5 & 1.19,0.24 & C3\\ 
1996.34 & 22.2 & 221,33.1 &  \multicolumn{1}{c}{\hspace{3mm}--} & \multicolumn{1}{c}{--} & 0.08,0.02 & Core\\ 
& &  35, 5.3 &  0.43,0.06 &  9, 7 & 0.24,0.05 & C9\\ 
& &  10, 1.5 &  0.97,0.09 & 12, 5 & 0.15,0.03 & C7\\ 
& &   6, 0.9 &  1.47,0.13 &  5, 5 & 0.20,0.04 & C6\\ 
1996.53 & 15.3 & 225,33.8 &  \multicolumn{1}{c}{\hspace{3mm}--} & \multicolumn{1}{c}{--} & 0.08,0.02 & Core\\ 
& &  17, 2.6 &  0.33,0.05 &  9, 8 & 0.08,0.02 & C9\\ 
& &  15, 2.2 &  0.75,0.09 &  4, 6 & 0.23,0.05 & C8\\ 
& &   4, 0.6 &  1.33,0.23 & 17, 9 & 0.43,0.09 & C7\\ 
& &   3, 0.4 &  4.12,0.34 & 15, 4 & 0.71,0.14 & C3\\ 
& &   1, 0.1 &  5.14,0.24 & 27, 2 & 0.12,0.02 & C2\\ 
1996.60 & 22.2 & 218,32.7 &  \multicolumn{1}{c}{\hspace{3mm}--} & \multicolumn{1}{c}{--} & 0.06,0.01 & Core\\ 
& &  17, 2.6 &  0.73,0.09 &  6, 6 & 0.26,0.05 & C8\\ 
& &   3, 0.4 &  1.16,0.19 & 15, 9 & 0.22,0.04 & C7\\ 
& &   4, 0.6 &  1.78,0.08 & 19, 2 & 0.05,0.01 & C6\\ 
1996.63 &  5.0 & 183,27.4 &  \multicolumn{1}{c}{\hspace{3mm}--} & \multicolumn{1}{c}{--} & 0.03,0.01 & Core\\ 
& &  22, 3.3 &  1.19,0.07 &  9, 3 & 0.24,0.05 & C7\\ 
& &   8, 1.2 &  4.09,0.27 & 20, 3 & 1.19,0.24 & C3\\ 
& &   7, 1.0 &  9.58,0.49 & 23, 2 & 3.39,0.68 & C1\\ 
& &   3, 0.4 & 12.06,0.44 & 10, 2 & 1.14,0.23 & X\\ 
1996.82 & 15.3 & 251,37.6 &  \multicolumn{1}{c}{\hspace{3mm}--} & \multicolumn{1}{c}{--} & 0.01,0.00 & Core\\ 
& &  13, 1.9 &  0.40,0.06 & 19, 7 & 0.08,0.02 & C9\\ 
& &  10, 1.5 &  1.49,0.16 & 12, 6 & 0.51,0.10 & C7\\ 
1996.90 & 22.2 & 260,39.0 &  \multicolumn{1}{c}{\hspace{3mm}--} & \multicolumn{1}{c}{--} & 0.04,0.01 & Core\\ 
& &  24, 3.6 &  0.22,0.04 & 22,10 & 0.08,0.02 & C10\\ 
& &   3, 0.4 &  0.76,0.12 & 16, 8 & 0.08,0.02 & C8\\ 
& &   7, 1.0 &  1.54,0.16 &  7, 6 & 0.38,0.08 & C7\\ 
& &   3, 0.4 &  4.69,0.29 & 13, 3 & 0.50,0.10 & C3\\ 
\hline
\multicolumn{7}{r}{continued on next Col.}\\
\end{tabular}
\end{minipage}
\end{table}
\setcounter{table}{1}
\begin{table}[htbp]
\caption{Modelfit parameters (continued from previous Col.).}
\begin{minipage}{\linewidth}
\renewcommand{\footnoterule}{}
\centering
\scriptsize
\begin{tabular}{l.d{3}d{2}d{2}d{2}l}
\hline
\multicolumn{1}{c}{Epoch} &
\multicolumn{1}{c}{$\nu$} &
\multicolumn{1}{c}{$S_{\rm peak}$} & 
\multicolumn{1}{c}{$r$} &
\multicolumn{1}{c}{$\phi$} &
\multicolumn{1}{c}{$\theta$} &
\multicolumn{1}{c}{Id.\footnote{Identification of the individual
components. If a component appeared only in a single epoch it is
labelled with X.}}\vspace{1pt}\\
 &
\multicolumn{1}{c}{[GHz]} &
\multicolumn{1}{c}{$[\frac{mJy}{beam}]$} & 
\multicolumn{1}{c}{[mas]} &
\multicolumn{1}{c}{[$^\circ$]} &
\multicolumn{1}{c}{[mas]} & \vspace{1pt}\\
\hline
1997.03 &  8.4 & 194,29.2 &  \multicolumn{1}{c}{\hspace{3mm}--} & \multicolumn{1}{c}{--} & 0.09,0.02 & Core\\ 
& &  22, 3.3 &  0.77,0.07 & 12, 5 & 0.23,0.05 & C8\\ 
1997.58 & 22.2 & 929,139.4 &  \multicolumn{1}{c}{\hspace{3mm}--} & \multicolumn{1}{c}{--} & 0.07,0.01 & Core\\ 
& &   6, 0.9 &  0.24,0.14 & 43,29 & 0.01,0.00 & C10\\ 
& &  32, 4.8 &  0.51,0.07 &  7, 7 & 0.32,0.06 & C9\\ 
& &   7, 1.0 &  1.69,0.07 & 11, 2 & 0.06,0.01 & C7\\ 
1997.93 &  8.4 & 385,57.7 &  \multicolumn{1}{c}{\hspace{3mm}--} & \multicolumn{1}{c}{--} & 0.10,0.02 & Core\\ 
& &  19, 2.8 &  0.72,0.06 & 14, 5 & 0.16,0.03 & C9\\ 
& &  15, 2.2 &  1.54,0.10 &  7, 3 & 0.30,0.06 & C7\\ 
& &   7, 1.0 &  3.18,0.24 & 15, 4 & 0.84,0.17 & C5\\ 
1999.41 &  8.4 & 922,138.3 &  \multicolumn{1}{c}{\hspace{3mm}--} & \multicolumn{1}{c}{--} & 0.10,0.02 & Core\\ 
& &  60, 9.0 &  0.68,0.07 & 10, 5 & 0.10,0.02 & C10\\ 
& &  17, 2.6 &  1.58,0.08 & 13, 2 & 0.20,0.04 & C8\\ 
& &   4, 0.6 &  3.16,0.19 &  6, 3 & 0.28,0.06 & C6\\ 
& &   4, 0.6 &  3.78,0.22 & 15, 3 & 0.40,0.08 & C5\\ 
1999.55 & 15.3 & 1157,173.6 &  \multicolumn{1}{c}{\hspace{3mm}--} & \multicolumn{1}{c}{--} & 0.06,0.01 & Core\\ 
& &  54, 8.1 &  0.40,0.03 & 15, 3 & 0.08,0.02 & C11\\ 
& &  26, 3.9 &  1.12,0.09 & 12, 4 & 0.39,0.08 & C9\\ 
& &   2, 0.3 &  1.89,0.26 & 10, 8 & 0.28,0.06 & C8\\ 
& &  10, 1.5 &  2.90,0.24 &  9, 4 & 1.17,0.23 & C6\\ 
1999.89 &  5.0 & 558,83.7 &  \multicolumn{1}{c}{\hspace{3mm}--} & \multicolumn{1}{c}{--} & 0.18,0.04 & Core\\ 
& &  39, 5.9 &  1.29,0.11 &  8, 5 & 0.13,0.03 & C9\\ 
& &  19, 2.8 &  2.64,0.15 & 10, 3 & 0.83,0.17 & C7\\ 
& &   6, 0.9 &  4.43,0.29 & 12, 3 & 1.00,0.20 & C4\\ 
& &   1, 0.1 &  6.92,0.98 & 18, 8 & 1.93,0.39 & C2\\ 
& &   7, 1.0 & 11.86,0.58 & 17, 2 & 4.79,0.96 & C1\\ 
2000.82 &  5.0 & 520,78.1 &  \multicolumn{1}{c}{\hspace{3mm}--} & \multicolumn{1}{c}{--} & 0.12,0.02 & Core\\ 
& &  35, 5.3 &  0.77,0.12 & 20, 8 & 0.10,0.02 & C11\\ 
& &  25, 3.9 &  1.63,0.09 & 12, 3 & 0.42,0.08 & C9\\ 
& &  11, 1.6 &  2.71,0.18 & 11, 3 & 0.74,0.15 & C7\\ 
& &   9, 1.4 &  4.16,0.26 & 11, 3 & 1.25,0.25 & C5\\ 
& &   8, 1.3 & 11.16,0.51 & 17, 2 & 4.43,0.89 & C1\\ 
2001.17 & 15.3 & 549,82.4 &  \multicolumn{1}{c}{\hspace{3mm}--} & \multicolumn{1}{c}{--} & 0.06,0.01 & Core\\ 
& &  72,10.8 &  0.45,0.04 & 20, 4 & 0.18,0.04 & C12\\ 
& &  14, 2.1 &  1.81,0.16 & 17, 5 & 0.70,0.14 & C9\\ 
& &   5, 0.8 &  3.06,0.22 & 13, 4 & 0.49,0.10 & C7\\ 
& &   3, 0.4 &  5.08,0.29 & 16, 3 & 0.51,0.10 & C4\\ 
\hline
\end{tabular}
\end{minipage}
\end{table}

\end{document}